# Optimal Placement of Public Electric Vehicle Charging Stations Using Deep Reinforcement Learning

Shankar Padmanabhan*
*Department of Mathematics*
*University of Texas at Austin*
Austin, USA
shankar.padmanabhan03@gmail.com

Aidan Petratos*
*Department of Computer Science*
*University of Texas at Austin*
Austin, USA
aiden.petratos@gmail.com

Allen Ting*
*Department of Computer Science*
*University of Texas at Austin*
Austin, USA
allenkting1@gmail.com

Kristina Zhou
*Department of Computer Science*
*University of Texas at Austin*
Austin, USA
zhou.kristina@utexas.edu

Dylan Hageman
*Department of Computer Science*
*University of Texas at Austin*
Austin, USA
dylanrhageman@utexas.edu

Jesse R. Pisel
*Department of Computer Science*
*University of Colorado, Boulder*
Boulder, USA
Jesse.Pisel@colorado.edu

Michael J. Pyrcz
*Cockrell School of Engineering,*
*Jackson School of Geosciences*
*University of Texas at Austin*
Austin, USA
mpyrcz@austin.utexas.edu

* Denotes equal authorship

I. *Abstract—* As climate change becomes a pressing issue, encouraging the use of electric vehicles (EVs) has become a popular solution to address the pollution associated with fossil fuel automobiles. Placing charging stations in areas with developing charging infrastructure is a critical component of the accessibility and future success of electric vehicles. In this work, we implement a supervised learning-based demand prediction model along with a Deep Q-Network (DQN) reinforcement learning (RL) algorithm to select optimal locations for charging stations. We use a gradient boosting model to predict EV charging demand at existing charging stations based on traffic data and points of interest (POIs) for select counties in the state of New York. The model outperforms other supervised learning methods with an $R^2$ score of 0.87, with the main predictors of demand being the traffic density near a station and the number of EV registrations in the surrounding ZIP code. We implement the RL environment by dividing Albany County, New York into a grid where each grid cell is assigned a demand value from the gradient boosting model. The DQN model trains to place ten charging stations within this grid and is rewarded based on the demand alleviated and area covered with each placement. Evaluations of the DQN agent's performance show that the agent achieves rewards that are, on average, 20-30% better than an agent performing random actions.

## II. INTRODUCTION

Electric vehicles (EVs) are essential to mitigating climate change, as they reduce carbon emissions and harmful pollutants that are often emitted from conventional vehicles (EERE, 2021). For instance, it is expected that EVs can reduce $CO_2$ emissions by 50 to 100 g of CO2 per kilometer (Nanaki and Koroneos, 2016). In recent years, the EV market has seen tremendous growth in the United States, with EV sales increasing from 100,000 vehicles in 2014 to over 350,000 vehicles in 2018 (AFDC, 2020). As a result of the emergence of EVs in the U.S., the demand for EV charging stations is also growing at a rapid pace. According to Atlas Public Policy's EValuateNY, a New York public database containing EV charging data, charging station usage as well as the number of EVs on the road has greatly increased in recent years. In New York County, the total energy consumption at charging stations funded by the New York State Energy Research and Development Authority (NYSERDA) has more than doubled from 4,425 kWh in the first quarter of 2014 to 11,036 kWh in the first quarter of 2019 (Atlas Public Policy, 2021).

As the number of EVs increases, there are concerns about the accessibility and convenience of charging stations for consumers (Bonges and Lusk, 2016). In the U.S., the number of currently installed Level 2 charging stations is only 12.0% of the total number of Level 2 stations needed to meet the projected charging demand in 2030 (Brown et al., 2020). Furthermore, many EV charging stations are brand specific. Thus, if the current charging infrastructure is not continually developed to meet demand or does not account for shifting demand patterns such as increases in population and EV ownership, the nation will have an inadequate charging network. A lack of available charging stations has also been shown to be directly related to consumers' decisions to buy EVs (Bonges and Lusk, 2016). Failing to address this problem would discourage consumers from switching to EVs. Therefore, there is an urgent need to maximize the coverage and efficiency of charging stations to effectively meet overall charging demand.

New charging stations should be installed in areas with high demand but optimizing charger placement is complex. Previous research has examined a multitude of EV charger characteristics. For example, Wagner et al. (2013) used regression analysis to analyze how points of interest (POIs) affect charging use. POIs are urban infrastructure buildings and areas (e.g., restaurants, stores, parks, etc.) that represent potential trip destinations for EVs. To determine optimal charging locations, various charging sites are ranked based on their attractiveness given surrounding POIs. Gopalakrishnan et al. (2016) used canonical correlation analysis to model the relationship between demand and external factors, such as nearby POIs and traffic densities, and predicted the hourly charging demand for each of 252 charging points in Northeast England. The researchers formulated the optimization of charger placement as a mixed pack and cover problem (MPC) which is solved by analyzing the desirability of charging points based on cost, demand, and geographical coverage.

However, potential changes in demand data over time have not been integrated into models, and to date, no one has used reinforcement learning (RL) for optimizing EV charger placement. In this work, we simulate an environment in which an RL agent can explore and locate the best placements for EV chargers in a region. A self-learning agent is adept at complex decision-making and can adapt to changing data to alleviate charging demand in both the present and the future. By combining supervised learning and RL methods with EV data, we can identify candidate-charging sites in many different environments with varying levels of EV charging infrastructure.

### III. DATA COLLECTION AND ANALYSIS

To optimize the placement of EV chargers with reinforcement learning, we first train a supervised Gradient Boosting model to predict spatial charging demand (kilowatt-hours per month) using traffic volume and POI data.

Charging station data are from Albany County, Monroe County, and Ulster County in New York, USA. Public charging locations are from the U.S. Department of Energy's Alternative Fuels Data Center (US Department of Energy, 2021) while EValuateNY (Atlas Public Policy, 2021) provides analytical data for charging networks within specific counties in New York. Combining the two datasets yields a single database and charging stations with no information are removed.

Traffic volume data for segments of road throughout the state are from the New York Department of Transportation for the year 2019 (New York Department of Transportation, 2021). The dataset provides the Annual Average Daily Traffic (AADT) for each road, which serves as a measure of traffic density in the supervised learning model. Traffic density affects charging station usage as it directly relates to the flow of EVs on roads in the area. In conjunction with traffic density, the number of EV registrations for all New York ZIP codes from EValuateNY is an additional feature. Energy consumption data from 2019 from EValuateNY is a target feature for the supervised model. However, due to sparse usage data, only the aggregate energy consumption (kWh/month) is available for a ZIP code. Therefore, to scale the data down to individual stations, the AADT values of the roads surrounding each charger are used to proportionally distribute the total energy demand for a ZIP code among different chargers. This method of scaling for a single station $E_S$ is shown in Equation (1).

$$E_s = E_{total} \times \left(A_S \div \left(\sum_{i=1}^{n} \frac{A_i}{n}\right)\right) \quad (1)$$

where, $E_{total}$, measured in kilowatt-hours, is the total energy usage of all the chargers in the ZIP code, $A_S$ represents the AADT value of a single charger in the ZIP code, $A_i$ represents the AADT value of the *i*-th charger, and *n* is the total number of chargers in the specific ZIP code. Due to the presence of outliers, as shown in the boxplots of Figure 1, energy usage and traffic data were both normalized using the Quantile Transformer from Scikit-learn (Pedregosa et al., 2011) before training the supervised learning model.

Lastly, POI information throughout New York is from OpenStreetMap, utilizing the Geofabrik website (Geofabrik, 2021) and GeoPandas (Kelsey et al., 2020). Each POI is assigned to one of 13 categories as shown in Table 1. In our model, the frequencies of each POI category within a 500m radius of the charging station are provided as additional features. These frequencies are shown in Figure 2. POIs provide information about building types in an area, which can influence traffic and EV charging use. Previous work has also found POIs to significantly impact nearby charging station usage (Wagner et al., 2013). The complete process of data collection is represented in Figure 3.

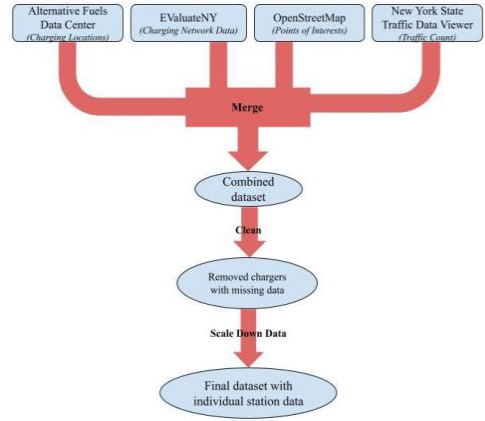

Figure 3. Process of data collection.

## IV. Demand Prediction Model

The supervised learning demand prediction model builds upon the work of Almaghrebi et al. (2020), which used supervised regression to predict charging demand for individual sessions. A set of features that EValuateNY uses to map trends with EV usage was selected to be used as input, as shown in Table 1. With these inputs, we evaluate different supervised learning algorithms proposed in Almaghrebi et al. (2020), which include the Random Forest, Gradient Boosting, Bagging, and Support Vector Machine models provided by the Scikit-learn library, and the XGBoost method taken from Chen and Guestrin (2016). These methods predict the energy demand for charging stations in three select counties across the state of New York that all possess a relatively large number of chargers. Hyperparameters for these models were tuned with Optuna (Akiba et al., 2019) using the default Bayesian optimization method, Tree Parzen Estimator, to discover the best values to input to achieve the highest R-squared score. Performance results and tuned hyperparameters for the different regression models are reported in Table 2.

| Regression | Performance (R-squared score) |
|---|---|
| Gradient Boosting | 0.868 |
| Random Forest | 0.336 |
| Bagging | 0.795 |
| XGBoost | 0.653 |
| Epsilon-Support Vector Regression | 0.363 |

Table 2. Supervised demand prediction model test results.

The Gradient Boosting Regressor performed the best at predicting demand, and we use it to set up the environment for our RL agent.

## V. Reinforcement Learning Framework

### A. Grid Representation

A 26 cell by 26 cell 2D grid model represents the charging station data in a selected subsection of Albany County shown on the map in Figure 4. Spatially, one grid cell is 250 by 250 meters ($m$), representing 62,500 square meters ($m^2$), which spans a reasonable walking distance from a central charging station to nearby POIs, sensu Wagner et al. (2013). Each grid cell contains the predicted demand value that results from placing a new charger in the center of that cell as calculated by the gradient boosting model. This grid is shown as a heatmap in Figure 5, showing the hot spots of demand that the agent should preferentially choose. The RL algorithm receives feedback for placing new stations in certain areas through a reward function.

### B. Model Framework

Our model utilizes Deep Q-Learning and the DQN (Deep Q-Network) algorithm to optimize the placement of charging stations. We use OpenAI Gym (Brockman et al., 2016) to create a training environment for the DQN agent. Given a demand grid of size $N \times N$, the discrete action space is defined by the set

$$a_t \in \{1, 2, \ldots, N \times N\} \quad (2)$$

where $a_t$ is the action taken at time $t$, and each action corresponds to placing a charger in one grid cell. The total number of chargers that an agent can place in the grid is restricted to 10 throughout the experiments.

The reward is calculated as the sum of the expected demand of a hypothetical charger in a grid cell as well as the area covered by the charger. We define coverage as a radius $r$, which is a large square with a side length of $2r + 1$ centered around the grid cell that the station is placed in (Figure 6). Each charging station placed by the agent covers new grid cells only if those cells are not within the radius of another charging station. Consequently, spreading out charging station placements leads to higher coverage rewards. To prevent the agent from maximizing reward based solely on charger coverage, we weight both coverage and demand in our reward function as follows. The predicted demand and coverage are both scaled using min-max normalization. Equation (3) represents the reward given at state $s_t$ for performing action $a_t$

$$R(s_t, a_t) = W_D D_{a_t} + W_C C \quad (3)$$

where $D_{a_t}$ represents the expected demand alleviated from taking action $a_t$, $C$ represents the new coverage gained by the new station placement, and $W_D$ and $W_C$ are the weights for demand and coverage, respectively, which are model hyperparameters. The agent's observation at time $t$ consists of two values: the difference between the agent's current accumulated demand and the maximum possible demand, and the difference between the current coverage and the maximum possible coverage. We implement a DQN agent in Keras for Python with the agent using a Sequential Model as the deep neural network combined with a Boltzmann policy. The code is partially adapted from Nicholas Renotte (2021).

The Deep Q-Network tested consists of an input layer, two hidden layers with ReLU activation functions with 300 and 256 neurons respectively, and an output layer using a linear activation function with 676 neurons, each corresponding to the Q-value associated with a charger placement in a grid cell. The hyperparameters for the baseline model are shown in Table 3. The DQN agent's rewards typically converged around 3,000 to 5,000 episodes, so we chose 5,000 episodes as the number of training episodes for analysis of the final models.

To examine the relationships between parameters for the DQN model, a manual search was conducted on various hyperparameters including the network architecture, activation functions, policy functions, demand radii, coverage radii, and weightings between demand and coverage. These parameters were implemented into the model, and the DQN agent was evaluated by comparing its episodic rewards to two benchmarks. The first we term the good reward, which is calculated by selecting a subset of 10 cells from the 30 highest demand cells that lead to the highest accumulated reward with any given set of parameters.

Since coverage also influences reward, the exact 10 highest demand cells will not always lead to the best possible reward when coverage and demand are weighted differently. Therefore, expanding the range of high demand cells and selecting a subset of the best cells ensures that the coverage portion of the reward is considered when establishing a value for what is considered a good reward. The second benchmark is a random agent, or an agent that randomly selects station locations in the grid. We call the ratio between our agent's reward and that of this random agent the random reward.

### C. Dynamic Demand

In a real-world scenario, placing a charging station would likely influence charging demand at nearby locations. While there is no data on this effect, we examine the possibility of changing demand by using Equation (4) to decrease demand at nearby grid cells and re-evaluate the DQN agent.

$$D_{S,new} = D_{S,old}\left(1 - \frac{1}{dist(S,C)+f}\right) \quad (4)$$

where $D_{S,new}$ and $D_{S,old}$ represent the new and old demand for grid cell $S$, respectively, $dist(S,C)$ represents the Cartesian distance from cell $S$ to the cell where the charger is placed ($C$), and $f$ represents the reciprocal of the factor by which the demand of the central cell is decreased. Equation (4) is applied in a series of trials with "dynamic demand". Two assumptions are made in these trials: $f$ is assigned a value of 2, and a radius $r_d$ is also set to 2, meaning that the equation is applied to a square of side length $2r_d + 1 = 5$ centered around the cell $C$.

## VI. RESULTS

Testing results from the supervised demand prediction model are shown in Table 2. Various models were tested and tuned using k-fold cross-validation with 5 folds, and the best performing model was the gradient boosting regressor with a R-squared score of about 0.87. The feature importances in Figure 7 document that traffic density and EV density are the most significant features, followed by certain POI categories such as education and sustenance.

Figure 7. Gradient Boosting Regressor Feature Importance.

For the RL algorithm, the mean episodic reward for the baseline network, shown in Figure 8, increased from -1.89 to 14.46 in 5,000 episodes, with cumulative rewards converging after about 3,000 episodes.

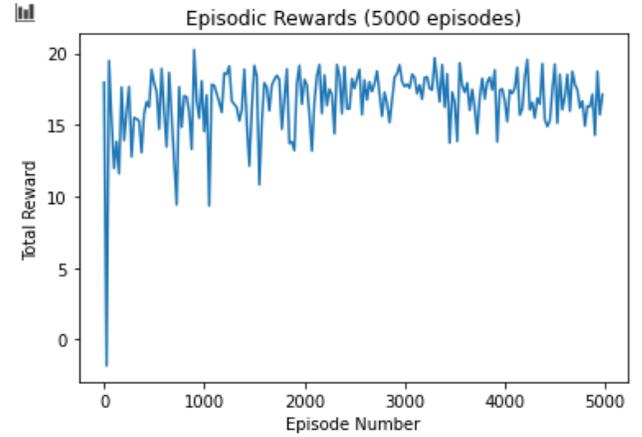

Figure 8. Mean Episodic Rewards for baseline DQN model training (Albany County)

Figure 9 shows two sets of actions taken by the agent during the training process in Albany County—one during episode 200, and one during episode 4,700, towards the end of training. Figure 10 shows the same sets of actions taken when training the agent on a grid of Monroe County. Black grid cells represent the charging station placements for the episode, and these actions are overlain with the charging demand heatmap. The final grid labeled Best Spots documents the grid cells with the ten highest demand values (note that the actual number of cells with these values can be more than ten, as the demand grid contains duplicates). Overall, the agent learns to place stations closer to the areas of highest demand but does not choose the ideal cells.

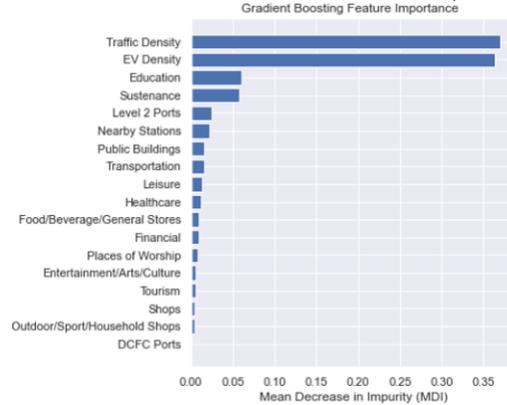

Figures 11 and 12 document the results of the evaluation of parameters for the DQN model with the good reward and random reward experiments, varying the coverage radius as well as the weights for coverage and demand. Additionally, in the trials with dynamic demand, the agent achieved lower rewards than in the trials where demand remained static across the grid.

## VII. DISCUSSION

### A. Demand Prediction Model

Given the mean R-squared score of 0.87 from the Gradient Boosting Regressor, the model predicted energy demand of Albany EV charging stations within a grid cell reasonably well. However, since the model heavily favors traffic and EV density as predictors of demand (Figure 7), POIs had much less of an impact than was originally hypothesized. When trained on just the POIs, the R-squared scores decrease for all of the supervised models tested. For Gradient Boosting in particular, the score dropped to a value of 0.43. This could be due to the way that POI data was collected, as 500m may not be the ideal radius to analyze nearby POIs. Alternatively, the specific POI makeup of Albany could lead the model towards favoring certain categories that may not apply to other regions.

While we consider the problem of predicting demand easy, the high feature importance of traffic density is a potential source of bias. Since traffic density was used to scale down energy usage, there may be some spurious autocorrelation between the two features. There are also outliers in demand and traffic density values at each station, shown by the boxplots in Figure 1. Without the Quantile Transformer, the mean R-squared scores decreased to about 0.6, suggesting that the high variance of input features had a notable effect on performance. By normalizing the data, the predictions from the gradient boosting model are on a consistent scale, which help in comparing results across different implementations of the DQN agent (e.g., different network architectures).

*B. Deep Reinforcement Learning*

When analyzing the specific actions taken by the agent, the model tends to place multiple charging stations in the same grid cell at the beginning. As training continues, the agent begins placing stations around the grid, moving towards cells with higher demand values. These results document that the model learns from its early shortcomings. However, the agent fails to find the grid cells with the highest demand, which might indicate that our approach must be altered before being applied to a larger region. In addition, the fact that many hyperparameters used within the model, such as the demand, coverage radii, and weightings, did not have a measurable impact on performance when adjusted might indicate that our grid is too homogeneous. In other words, the demand values of suboptimal grid cells may be too close to the values of the best grid cells for the agent to learn to prefer one over the other. This is supported by the fact that the random agent receives around 60-75% of the "good reward", so the average random station placement would already receive a significant percent of the reward.

The coverage radius metric presented mixed results: as the radius increased, the agent achieved lower percentages of the good reward, but its performance relative to the random agent improved (Figure 11a, b). We believe that this is because a random agent would perform worse with a higher coverage radius compared to the DQN agent. One possible explanation is that a random agent would get most of the possible coverage rewards and would need to rely on choosing high demand cells to further increase rewards, which is improbable for a random agent. This could also explain why our model's performance was closer to that of the random agent at higher coverage weightings (Figure 12a). For higher coverage weightings ($W_C > W_D$), our agent received a higher percentage of the good reward. This key takeaway made it clear that it is easier for the agent to learn to optimize coverage than demand (Figure 12c). When the demand is weighted higher than coverage ($W_D > W_C$), the DQN agent outperforms the random agent by greater margins as the demand weight increases (Figure 12b). However, the DQN agent's rewards also decreases relative to the good reward by greater amounts as the demand weight increases. This supports the conclusion that the DQN agent has more difficulty locating high demand areas, but still moves towards the optimal grid cells. The DQN agent achieved similar results when tested in Monroe, as can be seen in Figure 10, suggesting that the model could perform consistently across different regions.

Results fluctuated significantly in the trials with dynamic demand. Although there were trials in which the agent performed better than the random agent, the agent displayed a poor consistency. This could indicate that our agent failed to adapt its strategy adequately, instead performing well on a rare occasion by chance. While these results presented an interesting preview of more complex real-world situations, no solid conclusions can be drawn from these trials.

The results may be improved by further optimizing the environment parameters and the observation space, which currently consists of the total accumulated demand and the accumulated coverage. External sources of bias or outliers in the demand data are not believed to have skewed DQN training results. Based on the performance of the Gradient Boosting and DQN models, there are several notable correlations that can be utilized by future work to further improve RL environments for optimizing charging station placements.

VIII. CONCLUSION

Optimizing the placement of EV charging stations is a complex problem that must be solved to adequately support EV owners. Our solution to this problem utilizes a supervised gradient boosting model along with Deep Q-Learning (DQN). The supervised model predicts charging demand at stations with a R-squared score of 0.87, and the DQN model achieves rewards that are 20-30% better than a random agent.

Our research documents that supervised learning is an effective method for predicting charging demand at charging stations given multiple inputs. The best predictors of demand in the model were traffic density and the number of EVs in the same ZIP code as the station. However, energy use data for U.S. charging stations is sparse and difficult to find. Thus, having more detailed, centralized databases for charging station usage would greatly benefit future research in demand prediction. Experimental results from the DQN model indicate that DQN may not be the best solution for the problem of optimal charger placements. From our tests, we learned that coverage was easier for the agent to optimize for when compared to demand, which revealed the agent's difficulty in locating areas with higher demand. Since our framework only incorporates demand and coverage, other less complex methods may be better suited for this problem.

There are also many avenues of this research that could be improved upon in future work. The primary database used in this work was flawed in that its charging usage data was aggregated over a ZIP code, and better data at the individual station level would be ideal for future demand prediction. With the availability of data at the individual station level, the environment created for the RL agent could also be more precise with a continuous action space rather than using a grid. Charging wait times, which could determine the popularity of a charging station at a given spot, are also not a part of the models. Additionally, our model assumes a Level 2 charging station for every placement, but DC Fast Charging Stations can also be considered. Also, budget restrictions for varying station types or charging ports can be incorporated as well. Other RL algorithms can be tested in this environment, such as actor-critic, proximal policy optimization (PPO), or twin-delayed DDPG (TD3). Although there is still work to be done on the topic, this work provides a first pass examination of using reinforcement learning for optimizing charging infrastructure.

X. FIGURES

Table 1. Input features for demand prediction.

| Feature | Description |
|---|---|
| Traffic Density ($T_s$) | The average AADT of the 3 closest roads to the charging station |
| EV Density ($V_s$) | The number of EVs on the road in the same ZIP Code |
| Number of Nearby Stations ($N_s$) | The number of stations within a 1km radius of the station |
| Number of Level 2 Ports ($P_c$) | The number of Level 2 charging ports at the station |
| Number of DCFC Ports ($P_d$) | The number of DCFC charging ports at the station |
| Food/Beverage Shop ($I_1$) | POI category - frequency within 500m |
| Sustenance ($I_2$) | POI category - frequency within 500m |
| Outdoor/Sport Shop ($I_3$) | POI category - frequency within 500m |
| Entertainment ($I_4$) | POI category - frequency within 500m |
| Financial ($I_5$) | POI category - frequency within 500m |
| Education ($I_6$) | POI category - frequency within 500m |
| Healthcare ($I_7$) | POI category - frequency within 500m |
| Tourism ($I_8$) | POI category - frequency within 500m |
| Leisure ($I_9$) | POI category - frequency within 500m |
| Public Building ($I_{10}$) | POI category - frequency within 500m |
| General Shop ($I_{11}$) | POI category - frequency within 500m |
| Transportation ($I_{12}$) | POI category - frequency within 500m |
| Places of Worship ($I_{13}$) | POI category - frequency within 500m |

Table 2. Supervised demand prediction model test results and hyperparameters (extended).

| Regression | Parameters | Performance (R-squared score) |
|---|---|---|
| Gradient Boosting | learning_rate = 0.038<br>n_estimators = 6873<br>max_features = 0.894<br>max_depth = 14.444<br>criterion = Mean Squared Error<br>min_samples_split = 30<br>min_samples_leaf = 1 | 0.868 |
| Random Forest | n_estimators = 2943<br>max_features = 0.400<br>max_depth = 6.236<br>criterion = Mean Absolute Error<br>min_samples_split = 5<br>min_samples_leaf = 22 | 0.336 |
| Bagging | n_estimators = 4040<br>max_features = 0.943 | 0.795 |
| XGBoost | objective = Mean Squared Error<br>booster = 'gbtree'<br>n_estimators = 7880<br>learning_rate = 0.317<br>gamma = 0<br>max_depth = 14.01<br>min_child_weight = 38 | 0.653 |
| Epsilon-Support Vector | kernel = 'rbf'<br>C = 1.0 | 0.363 |
| Regression | | |

Table 3. DQN model default training hyperparameters.

| Parameters | Performance (R-squared score) |
|---|---|
| Policy | BoltzmannQPolicy |
| Experience Replay Memory Size | 50,000 |
| Total episodes | 5,000 |
| Target model update interval | 0.01 |
| Optimizer | Adam |
| Learning rate | 0.001 |
| Loss Function | Mean Squared Error (MSE) |
| Batch size | 32 |
| Discount Factor, $\gamma$ | 0.99 |
| Coverage Radius | 2 |
| Demand Weight, $W_D$ | 1 |
| Coverage Weight, $W_C$ | 1 |

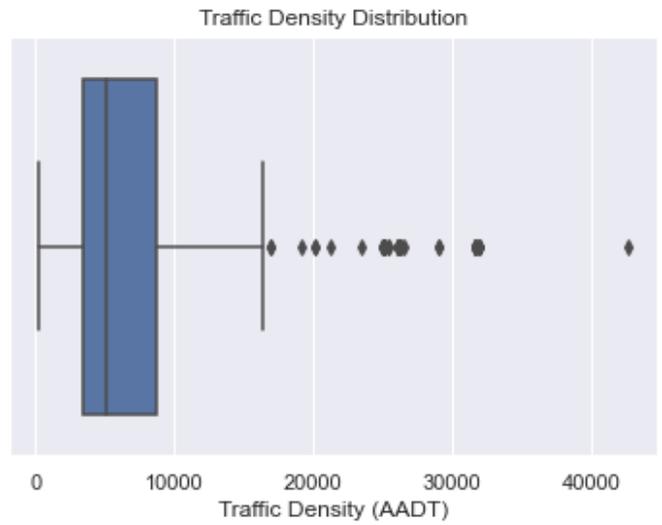

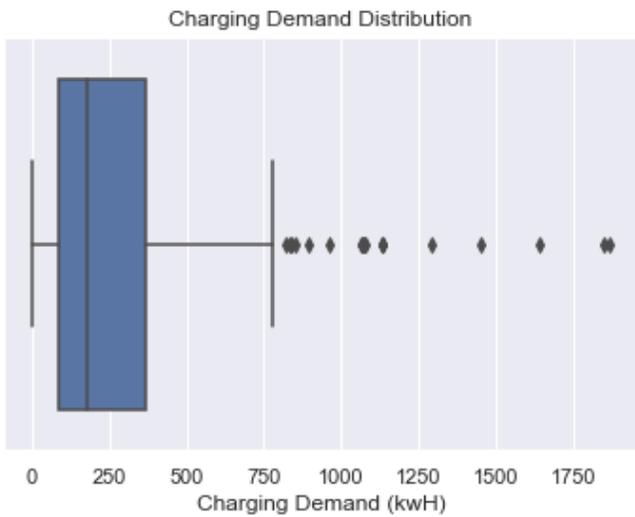

Figure 1. Boxplots of traffic density and energy demand across charging stations.

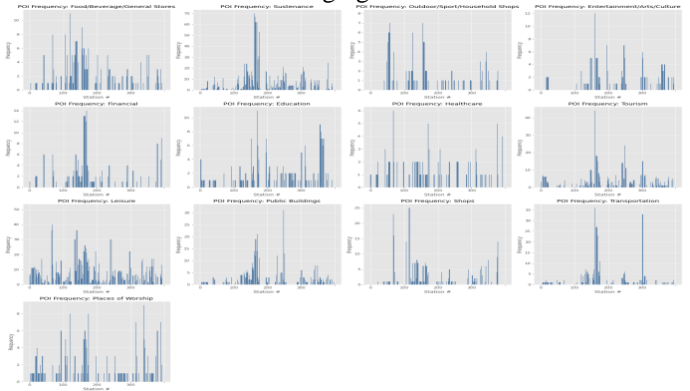

Figure 2. POI frequencies for charging stations in the dataset.

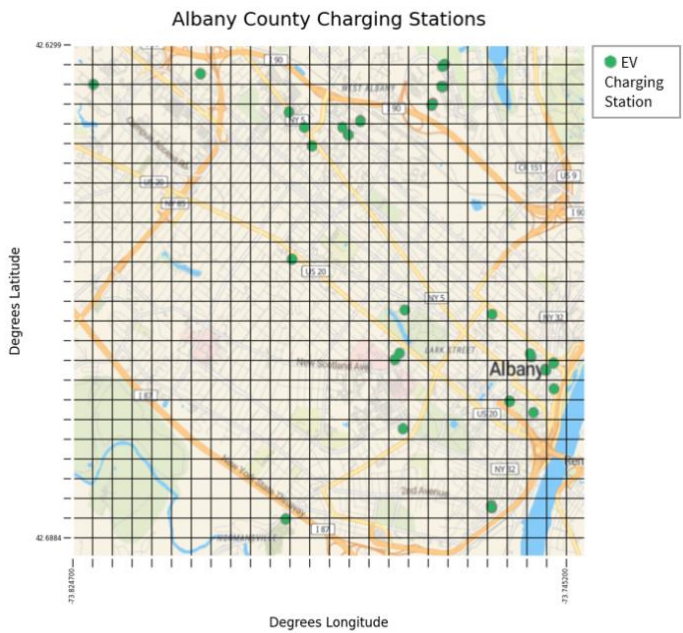

Figure 4. Subsection chosen in Albany County. (AFDC, 2021)

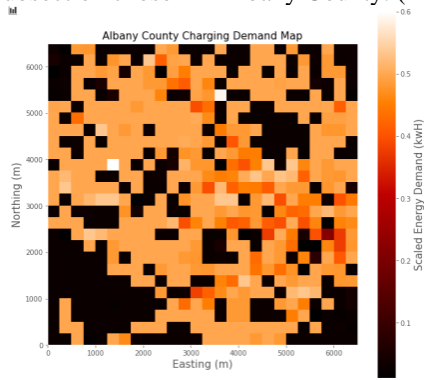

Figure 5. Charging Demand Heatmap.

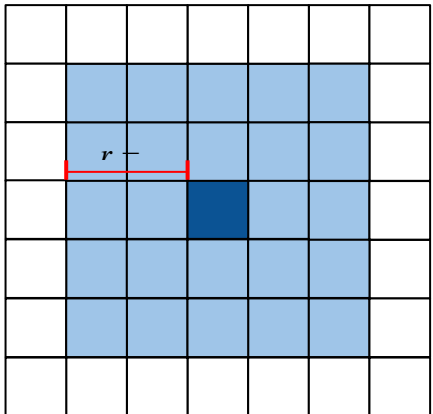

Figure 6. Definition of coverage radius. The dark blue square in the center represents the grid cell in which the charger is placed, and the light blue squares represent the area that is covered by the charger.

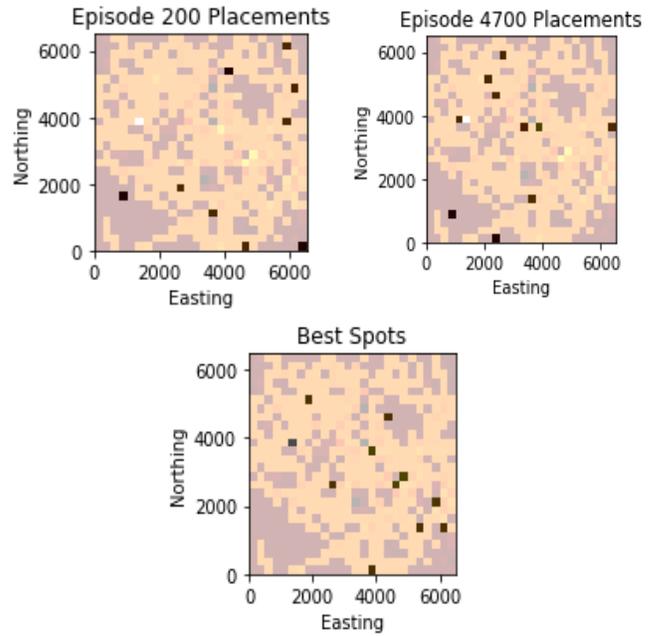

Figure 9. Actions taken during DQN training vs. Best Spots (Albany County).

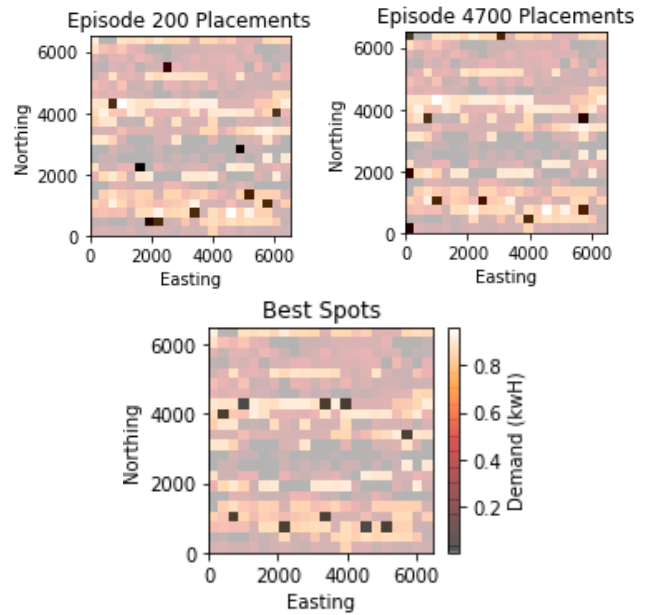

Figure 10. Actions taken during DQN training vs. Best Spots (Monroe County).

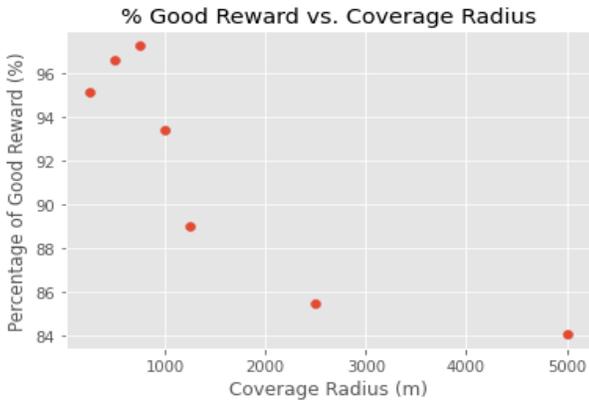
(a)

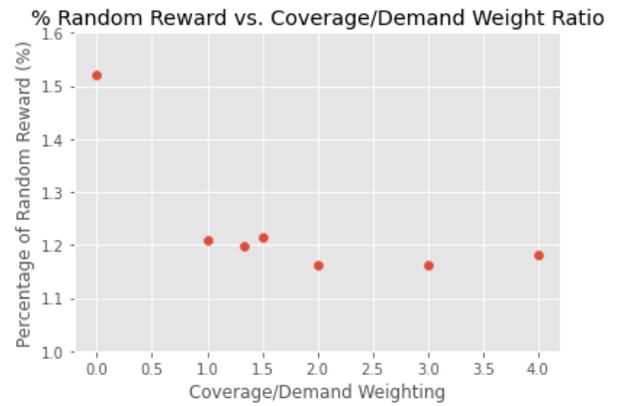
(b)

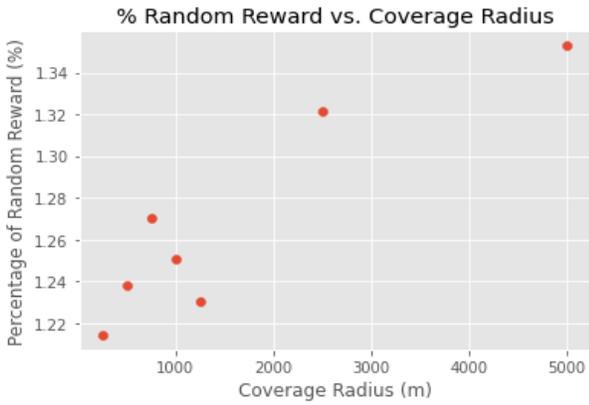
(b)

Figure 11. DQN evaluations compared to "good" rewards and random rewards. (a) Good Reward vs. Coverage Radius; (b) Random Reward vs. Coverage Radius

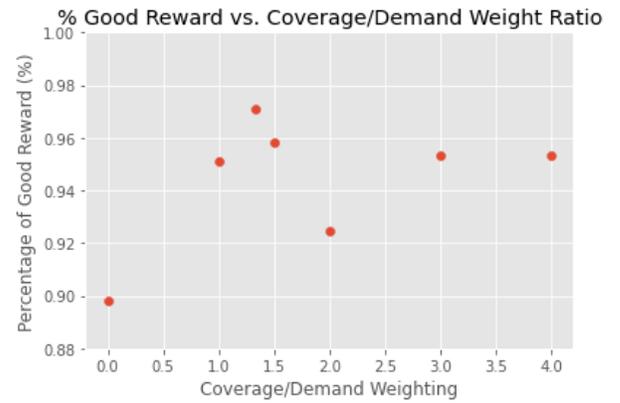
(c)

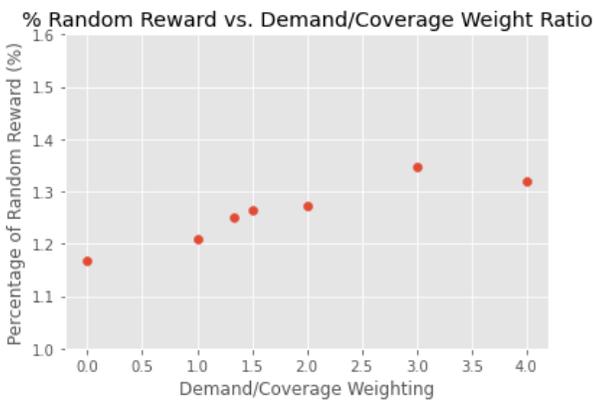
(a)

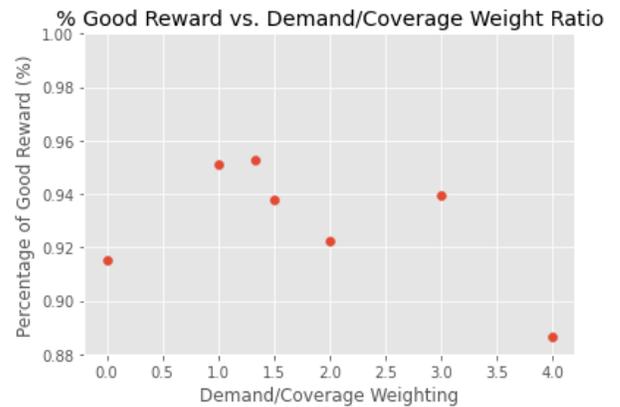
(d)

Figure 12. DQN evaluations compared to "good" rewards and random rewards (a) Random Reward vs. Increasing Coverage Weight; (b) Random Reward vs. Increasing Demand Weight; (c) Good Reward vs. Increasing Coverage Weight; (d) Good Reward vs. Increasing Demand Weight.